\documentclass[fleqn,10pt]{wlpeerj}

\usepackage{hyperref}
\usepackage[style=apa,backend=biber,apamaxprtauth=99]{biblatex}
\usepackage{soul}
\usepackage{svg}
\usepackage{pgf}
\usepackage{graphicx}
\usepackage{csquotes}
\usepackage{booktabs}
\usepackage{todonotes}
\usepackage{tabularx}

\usepackage{tikz}
\usetikzlibrary{shapes.geometric, arrows}

\tikzstyle{startstop} = [rectangle, rounded corners, minimum width=3cm, minimum height=1cm,text centered, text width=3cm, draw=black, fill=red!30]
\tikzstyle{io} = [trapezium, trapezium left angle=70, trapezium right angle=110, minimum width=3cm, minimum height=1cm, text centered, text width=3cm, draw=black, fill=blue!30]
\tikzstyle{process} = [rectangle, minimum width=3cm, minimum height=1cm, text centered, text width=3cm, draw=black, fill=orange!30]
\tikzstyle{decision} = [diamond, minimum width=3cm, minimum height=1cm, text centered, text width=3cm, draw=black, fill=green!30]
\tikzstyle{arrow} = [thick,->,>=stealth]
\tikzstyle{arrownotip} = [thick,-,>=stealth]

\usepackage{xparse}

\let\oldsection\section
\makeatletter
\newcounter{@secnumdepth}
\RenewDocumentCommand{\section}{s o m}{%
  \IfBooleanTF{#1}
    {\setcounter{@secnumdepth}{\value{secnumdepth}}
     \setcounter{secnumdepth}{0}
     \oldsection{#3}
     \setcounter{secnumdepth}{\value{@secnumdepth}}}
    {\IfValueTF{#2}
       {\oldsection[#2]{#3}}
       {\oldsection{#3}}}
}
\let\oldsubsection\subsection
\makeatletter
\RenewDocumentCommand{\subsection}{s o m}{%
  \IfBooleanTF{#1}
    {\setcounter{@secnumdepth}{\value{secnumdepth}}
     \setcounter{secnumdepth}{0}
     \oldsubsection{#3}
     \setcounter{secnumdepth}{\value{@secnumdepth}}}
    {\IfValueTF{#2}
       {\oldsubsection[#2]{#3}}
       {\oldsubsection{#3}}}
}
\let\oldparagraph\paragraph
\makeatletter
\RenewDocumentCommand{\paragraph}{s o m}{%
  \IfBooleanTF{#1}
    {\setcounter{@secnumdepth}{\value{secnumdepth}}
     \setcounter{secnumdepth}{0}
     \oldparagraph{#3}
     \setcounter{secnumdepth}{\value{@secnumdepth}}}
    {\IfValueTF{#2}
       {\oldparagraph[#2]{#3}}
       {\oldparagraph{#3}}}
}
\makeatother

\title{Don’t mention it: An approach to assess challenges to using software mentions 
for citation and discoverability research}

\author[1,2]{Stephan Druskat} 
\author[3]{Neil P. Chue Hong} 
\author [4]{Sammie Buzzard} 
\author [5]{Olexandr Konovalov} 
\author [5]{Patrick Kornek} 

\affil[1]{German Aerospace Center (DLR), Institute for Software Technology, Berlin, Germany}
\affil[2]{Humboldt-Universität zu Berlin, Department of Computer Science, Berlin, Germany}
\affil[3]{EPCC, University of Edinburgh, Edinburgh, United Kingdom}
\affil[4]{School of Earth and Environmental Sciences, Cardiff University, Cardiff, United Kingdom}
\affil[5]{School of Computer Science, University of St Andrews, St Andrews, United Kingdom}
\corrauthor[1]{Stephan Druskat}{stephan.druskat@dlr.de}

\begin{abstract}
Datasets collecting software mentions from scholarly publications can potentially be used 
for research into the software that has been used in the published research,
as well as into the practice of software citation.
Recently, new software mention datasets with different characteristics have been published.
We present an approach to assess the usability of such datasets for research on research software.
Our approach includes sampling and data preparation, manual annotation for quality and mention characteristics, and annotation analysis.
We applied it to two software mention datasets for evaluation based on qualitative observation.
Doing this, we were able to find challenges to working with the selected datasets to do research.
Main issues refer to the structure of the dataset, the quality of the extracted mentions ($54\%$ and $23\%$ of mentions respectively are not to software),
and software accessibility.
While one dataset does not provide links to mentioned software at all,
the other does so in a way that can impede quantitative research endeavors: (1) Links may come from different sources and each point to different software for the same mention. (2) The quality of the automatically retrieved links is generally poor (in our sample, $65.4\%$ link the wrong software). (3) Links exist only for a small subset (in our sample, $20.5\%$) of mentions, which may lead to skewed or disproportionate samples.
However, the greatest challenge and underlying issue in working with software mention datasets
is the still suboptimal practice of software citation:
Software should not be mentioned, it should be cited following the software citation principles.
\end{abstract}

\begin{document}

\flushbottom
\maketitle
\thispagestyle{empty}

\section*{Introduction}
\label{sec:introduction}

Until recently, a key challenge in trying to understand the research software landscape was a lack of knowledge of what software is being used.
While there are some discipline-specific software catalogues and lists, 
and recent initiatives such as the Netherlands eScience Center Research Software Directory~\parencite{spaaksResearchSoftwareDirectory2020a} and the Research Software Encyclopedia~\parencite{theresearchsoftwareencyclopediaprojectResearchSoftwareEncyclopedia2021a},
there is no single, comprehensive directory or curated list of research software
whose use is attested in the literature.

Approaches to build datasets that contain software references, e.g. for
specific disciplines, often take publications as a starting point and identify 
software that was mentioned in the publication, either manually~\parencite{duSoftciteDatasetDataset2021} or through machine learning~\parencite{wadeCORD19SoftwareMentions2021a,IstrateEtAl2022a,SchindlerEtAl2022a}.
Domain registries such as \textit{swMATH} for mathematical software \parencite{DalitzSperberChrapary2020}
and \textit{ASCL} for astrophysics software \parencite{ASCL-JORS} use these techniques to identify possible
candidates for inclusion.
Other approaches rely on automatically mining code repositories,
looking for key markers such as citation files or DOIs~\parencite{eitzenResearchSoftwarePublication2020}.

Researchers who want to do quantitative empirical research on research software and research software engineering (see~\textcite{FeldererEtAl2023})
have a need for datasets with specific features that allow them to access research software metadata and artifacts, including source code,
i.e., datasets need to include URLs to repositories that store the relevant metadata or artifacts.

\newpage

\noindent Examples for such research include
\begin{itemize}[noitemsep]
    \item accessing source code to study implementation details in research software;
    \item accessing source code repositories to study
    software metadata such as license information;
    \item accessing features of source code repositories such as
    the software's version history, issue trackers or integration processes (pull\slash merge requests)
    to study development processes and software provenance~\parencite{dlr147617}.
\end{itemize}
\noindent Recently, five datasets
have been released
that can 
potentially be used for 
the type of research described above.
The \textit{SoftCite} project~\parencite{duSoftciteDatasetDataset2021} is a human curated list of $4,093$ software mentions in the life and social sciences. This dataset, in turn, was used by a team at the Chan Zuckerberg Initiative to train a machine learning model that has been used to identify software references in the \textit{CORD-19} collection of COVID-19-related research papers~\parencite{wangCORD19COVID19Open2020b}, which has been published as a raw dataset: \textit{CORD-19 Software Mentions}~\parencite{wadeCORD19SoftwareMentions2021a} (\textit{CSM}).
The \textit{CZ Software Mentions} dataset~\parencite{IstrateEtAl2022b} (\textit{CZI}) provides software mentions from the biomedical literature, their sources, textual context and metadata, extracted by a
trained SciBERT model~\parencite{BeltagyEtAl2019,IstrateEtAl2022a}.
A subset of the complete corpus also disambiguates software entities, and provides links to software source code and/or artifact repositories.
\textit{SoftwareKG-PMC}~\parencite{SchindlerDavidEtAl2021} provides a knowledge graph of software mentions generated through a SciBert~\parencite{BeltagyEtAl2019} model trained on
a gold standard corpus of software mentions~\parencite{SchindlerEtAl2021,SchindlerEtAl2022a}.
\textcite{EscamillaEtAl2022a} extracted URLs for git hosting platforms from the ArXiv and PMC corpora (\textit{Extract-URLs},~\textcite{Escamilla_Extract-URLs}).

In this paper, we present our approach to assessing the usability of software mention datasets for
\begin{enumerate*}[label=(\arabic*)]
    \item quantitative research on research software that requires access to software metadata or artifacts (i.e. for mining software repositories), and
    \item research into the practice of software citation.  
\end{enumerate*}
We understand ``usability'' as the potential for a dataset to be used without further
processing to access the source code of the mentioned software that is recorded in the dataset.
An example for high usability would be the inclusion in a dataset of a correct URL
to the source code repository for each software in the dataset.
Our approach was originally developed to use CSM~\parencite{wadeCORD19SoftwareMentions2021a} in order to answer research questions (see~\textit{\nameref{subsec:methods:rqs}})
from the above types of research: 
\begin{enumerate*}[label=(\arabic*)]
    \item What is the impact of licenses on the type of software mention?
    \item Has the practice of software citation improved as compared to~\textcite{howisonSoftwareScientificLiterature2015}?
\end{enumerate*}
We consecutively applied our approach to assess the usability of CZI~\parencite{IstrateEtAl2022b},
which gave us opportunity to evaluate our approach,
and compare the usability of two datasets that have been created in a similar way (see~\textit{\nameref{subsec:methods:datasets}}).
The approach consists mainly of the manual annotation of a stratified sample of the dataset
with quality categories, categories of software mentions, and accessibility categories,
and the subsequent analysis of the annotations.




In the following sections, we present related work (\textit{\nameref{subsec:related-work}}) and
describe our assessment approach and
the methodology to apply this approach to the CSM and CZI datasets (\textit{\nameref{sec:methods}}).
We then present the results of the assessment and the results of
two
exploratory studies 
based on the assessed datasets (\textit{\nameref{sec:results}}).
Finally, we conclude with suggestions for future work (\textit{\nameref{sec:conclusions}}).

\subsection*{Related work}
\label{subsec:related-work}
In 2015, Howison and Bullard described challenges with identifying and finding software that has been used for biology research,
and crediting software authors~\parencite{howisonSoftwareScientificLiterature2015}.
These challenges relate to problematic practices of software citation:
Howison and Bullard found that of the publications that mentioned software in any way,
only $39\%$ cited a formal publication relating to the software in the references,
while informal mentions are included in $43\%$ of investigated publications.
Such informal mentions often fail to provide credit to software authors.
Furthermore, only in $28\%$ of cases were they able to identify the software versions that had been used for the research presented in the respective publications,
and only $5\%$ of the respective versions were referenced in a way that made it possible to find the actual software versions.

Since the publication of Howison and Bullard's paper,
work has been done in the scholarly communications and research software communities
to address the state of software citation practice.
\textcite{smithSoftwareCitationPrinciples2016} define the principles of software citation.
These principles (italicised, see~\textcite{smithSoftwareCitationPrinciples2016}
for their definitions) address the challenges to software identification, findability, and creditability mentioned by \textcite{howisonSoftwareScientificLiterature2015}, when applied in practice: 

\begin{itemize}[nosep]
    \item \textit{Importance} and \textit{unique identification} disallow informal mentions that may obscure the identity of the software.
    \item \textit{Persistence}, \textit{accessibility} and \textit{specificity} enable findability of and access to the specific version of referenced software.
    \item \textit{Credit and attribution} enable credit for software authors.
\end{itemize}

The FORCE11 Software Citation Implementation Working Group (2017--2023)\footnote{\url{https://www.force11.org/group/software-citation-implementation-working-group}. SD and OK were members of the working group that was co-chaired by NCH.} worked with different stakeholders to endorse the software citation principles, helped implementing them, and developed guidelines for different stakeholder groups, e.g., for software developers~\parencite{chuehongSoftwareCitationChecklist2019}, journal authors~\parencite{chuehongSoftwareCitationChecklist2019a}, publishers~\parencite{katzRecognizingValueSoftware2021}, software registries~\parencite{registriestaskforceonbestpracticesforsoftwareNineBestPractices2020}, and libraries~\parencite{schmidtbirgitLIBER2021Session2021}.

Software citation practices based on the software citation principles are also implemented in, or supported by: 
software metadata formats, that enable software authors to provide correct and complete citation metadata (e.g. Citation File Format~\parencite{druskatstephanCitationFileFormat2021b},
CodeMeta~\parencite{jonesCodeMetaExchangeSchema2017b});
open access repositories and archives that provide DOIs or other unique identifiers for software 
(e.g. Zenodo~\parencite{zenodo}, Figshare, Software Heritage~\parencite{cacm-2018-software-heritage});
source code platforms that display citation information for software (versions)
(e.g. GitHub~\parencite{EnhancedSupportCitations2021});
reference managers that support import of correct and complete software citation metadata
(e.g. Zotero, JabRef);
publications that support and recommend citation of software versions~\parencite{katzRecognizingValueSoftware2021}.

Despite these activities towards better software citation, and an increase in the number of DOI registrations for software~\parencite{fennerDOIRegistrationsSoftware2018a},
good software citation practice, it seems, has not yet permeated research culture:
only ``$3.24\%$ of all software DOIs registered by Zenodo are traceably cited at least once''~\parencite[p. 2]{vandesandtPracticeMeetsPrinciple2019}.
Instead, informal mentions of software in publications are still commonplace,
which makes it hard to track software usage in research~\parencite[see][]{duSoftciteDatasetDataset2021}.
This in turn is not only disadvantageous for the sustainability of the software in question \parencite[see][for a brief discussion of the relationship between citation and research software sustainability]{druskatResearchSoftwareSustainability2021b}, it also impedes quantitative research on different aspects of research software:
instead of being able to work with citation graphs that include software directly~\parencite{druskatSoftwareDependenciesResearch2020,PeroniShotton2020},
researchers are forced to trace software mentions.
This is often a time-consuming and costly process: to build the \textit{SoftCite} dataset,
36 student assistants were paid for 2 years
to annotate software mentions in full-text PDF files~\parencite[p. 872]{duSoftciteDatasetDataset2021}.
An alternative to the manual extraction of software mentions is the application of machine learning to datasets of literature.
In this paper, we report on two datasets that were created using machine learning models.

\section*{Methods}
\label{sec:methods}

The availability of new, large software mention datasets (see \textit{\nameref{sec:introduction}})
promises improved access to research software whose use has been reported in the literature.
If these datasets include direct links to software metadata and\slash or artifacts for research software,
they can be used to build corpora of research software source code, or extract samples for empirical research.
Additionally, these datasets may potentially be used to analyse the practice of software citation.
We iteratively developed an approach to assess
the usability of software mention datasets for these purposes.
We then applied the approach to two of the datasets mentioned above.
Our approach included three steps:
\begin{enumerate}
    \item Take a stratified proportionate random sample from a dataset and prepare the sample for annotation.
    \item Manually annotate the sample for mention extraction quality, mention categories and quality, following a set of annotation guidelines.
    \item Analyse the sampling and preparation workflows, as well as the annotations, to assess the usability of the dataset, and report preliminary results for our research questions where possible.
\end{enumerate}
Code, samples and annotated data are available as~~\textcite{druskatDonMentionIt2021}.

\subsection*{Research questions}
\label{subsec:methods:rqs}

We developed and applied our approach to answer the following research questions.

\textbf{RQ1: Are software mention datasets usable as data sources for research on research software?}\label{para:rq1}
More precisely:
\textbf{RQ1.1: Are software mention datasets usable as data sources for research on research software that requires access to software metadata or artifacts?}\label{para:rq1-1}
and
\textbf{RQ1.2: Are software mention datasets usable as data sources for research into the practice of software citation?}\label{para:rq1-2}

The goal of these research questions is to identify features of datasets that allow their use for research on research software, 
and vice versa, identify features that make datasets' use for this purpose harder or impossible.
Intuitively, the inclusion in a dataset of links to software artifacts or metadata would enable research.
Results would provide evidence of additional features whose presence, absence, or characteristics can enable or preclude reuse for research purposes.

\textbf{RQ2: Is open source software more cited in a way that allows credit for software authors than closed source software?}\label{para:rq2}

The goal of this research question is
to understand if the way that the software is licensed has an impact on the way that the software is cited or mentioned in publications. 
\cite{howisonSoftwareScientificLiterature2015} define seven types of software mentions in publications, categorised as formal citations (to publications, user manuals or project websites), explicit mentions (like an instrument, of a URL, or just the software name) or implicit mentions (not even a name). We hypothesise that commercial software is cited more frequently using an in-text name mention or citation to project name or website, but that open source software is cited more frequently with a repository or associated research publication in the reference. The latter makes it easier to credit the authors. Results would provide evidence concerning whether the increasing prevalence of Open Science / Open Research approaches could improve the quality of software citation.

\textbf{RQ3: Has the practice of software citation represented in software mention datasets improved in comparison to the practice described in~\textcite{howisonSoftwareScientificLiterature2015}?}\label{para:rq3}

The goal of this research question is
to find out if the practice of software citation has improved
since the publication of~\textcite{howisonSoftwareScientificLiterature2015}.
Since 2015, a number of contributions have been made
towards better software citation.
In 2016 the software citation principles were published~\parencite{smithSoftwareCitationPrinciples2016}.
They describe how software should be cited, and what metadata
the respective references should ideally include:
The software itself should be cited, rather than a
substitute output, e.g., a paper describing software.
The reference should also name the authors to enable academic credit,
include a persistent identifier to the persisted artifact of the exact version of the software that was used, and allow access to the software
itself.
Following the principles paper, which was the main output
of the FORCE11 Software Citation Working Group\footnote{\url{https://force11.org/group/software-citation-working-group/}}, the follow-up Software Citation Implementation WG\footnote{\url{https://force11.org/groups/software-citation-implementation-working-group/}} (2017--2023) created a number of outputs
addressing different stakeholders to improve software citation practice:
software developers~\parencite{chuehongSoftwareCitationChecklist2019}, authors~\parencite{chuehongSoftwareCitationChecklist2019a}, publishers~\textcite{jaySoftwareMustBe2021} and libraries~\parencite{chuehongRecognisingValueSoftware2021a}.
Additionally, the FAIR Principles for Research Software have drawn attention to the importance of software metadata that also ``enables and encourages the citation of software''~\parencite[8]{FAIR4RS_2021}.
And metadata formats were developed to make it easier
to provide correct and complete software citation metadata in the
first place~\parencite{druskatstephanCitationFileFormat2021b,jonesCodeMetaExchangeSchema2017b}.

While we cannot prove causation between any of the
above-mentioned activities and outputs and the data in software mention datasets, we hypothesise that software citation practice will overall have
improved since the publication of~\textcite{howisonSoftwareScientificLiterature2015}.
Specifically, we expect that those of ~\citeauthor{howisonSoftwareScientificLiterature2015}'s (\citeyear{howisonSoftwareScientificLiterature2015}) categories of mentions that
reflect the principles better are found relatively more often in 
mentions from publications in the datasets that were published in or after 2016.

\subsection*{Software mention datasets}
\label{subsec:methods:datasets}
We applied our approach to two datasets:
\begin{itemize}
    \item \textit{CORD-19 Software Mentions}~\parencite{wadeCORD19SoftwareMentions2021a} (\textit{CSM})
    \item \textit{CZ Software Mentions: A large dataset of software mentions in the biomedical literature}~\parencite{IstrateEtAl2022b} (\textit{CZI})
\end{itemize}

\noindent CSM contains lists of software mentioned in $77,000$ papers, and exceeds the size of the dataset used by
\textcite[random sample of $90$ papers]{howisonSoftwareScientificLiterature2015} by a
factor of over $850$. It was created by detecting software mentions in the 
CORD-19~\parencite{wangCORD19COVID19Open2020b} collection of full-text research papers related to the SARS‑CoV‑2 virus,
using a machine learning model originally trained and evaluated on the \textit{SoftCite} dataset~\parencite{duSoftciteDatasetDataset2021}.

CZI contains software mentioned in $\approx20.7$ million papers~\parencite{IstrateEtAl2022a}, thereby exceeding the size of CSM by a factor of $\approx26,000$.
It was created by detecting software mentions in the NIH PMC-OA Commercial and Non-Commercial subsets, and a custom collection of papers provided by various publishers
to the Chan Zuckerberg Initiative. The detection used a machine leraning model based on SciBERT\parencite{BeltagyEtAl2019}, also trained on the \textit{SoftCite} dataset~\parencite{duSoftciteDatasetDataset2021}.

The datasets were chosen for different reasons. Firstly, both datasets have likely been prepared using the same or different versions of the same machine learning model\footnote{See the code at~\textcite{Veytsman2022} which uses the model that was used to create CSM (see~\textcite{TheSoftwareMentionExtractionauthors2022}).}, share authors (I. Williams), and have been prepared at the same institution (Chan Zuckerberg Initiative, Redwood City, CA, USA). This suggested that the methods applied in dataset creation would not have to be factored into a comparison of the two datasets. 

Secondly, we were involved in preliminary work on CSM, which led to an earlier version of this paper. This preliminary work was conducted during a hack event at the \textit{Software Sustainability Institute}'s
\textit{Collaborations Workshop 2021 Hack Day}\footnote{\url{https://software.ac.uk/cw21/hack-day}}~\parencite{KonovalovEtAl2021}. During the hack event, a random sample of 100 mentions from CSM was taken and manually cleaned, and annotated with categories pertaining to the quality of the mention with respect to accessibility of the software and the quality of the mention extraction. Additionally,
for each mention in the sample, identification of a source code repository was attempted, and the respective URL added to the dataset. The resulting dataset (\texttt{CORD19\_software\_popularity\_sampled\_QA\_DOI.csv}) is available as part of~\textcite{druskatDonMentionIt2021}.

\subsection*{Sampling}
\label{subsec:methods:sampling}

Usually, sampling is a means to the end of obtaining data for a study, e.g. in sample studies along the lines of~\textcite[p. 16f.]{StolFitzgerald2018}.
In our case, the sampling process was part of the study to answer RQ1, if software mention datasets are usable as data sources for research on research software.
Here, we describe our methods for sampling CSM and CZI.
We discuss results towards answering RQ1 (RQ1.1, RQ1.2) in \textit{\nameref{sec:results}}.

To conduct our study towards answering RQ2 and RQ3, we took samples of both CSM and CZI.
Sampling was done in Jupyter Notebooks~\parencite{jupyter, Kluyver2016jupyter} using NumPy~\parencite{numpy, 2020NumPy-Array}, Pandas~\parencite{Thepandasdevelopmentteam2023}, Dask~\parencite{TheDask2023.3.1developers2023}, Matplotlib~\parencite{matplotlib, Hunter:2007}, SciPy~\parencite{SciPy-NMeth,TheSciPy1.10.1developers2021} and scikit-learn~\parencite{sckikit}.

\begin{figure}[h!t]
    \centering
    \begin{minipage}{0.45\textwidth}
        \centering
        \includegraphics[width=\textwidth]{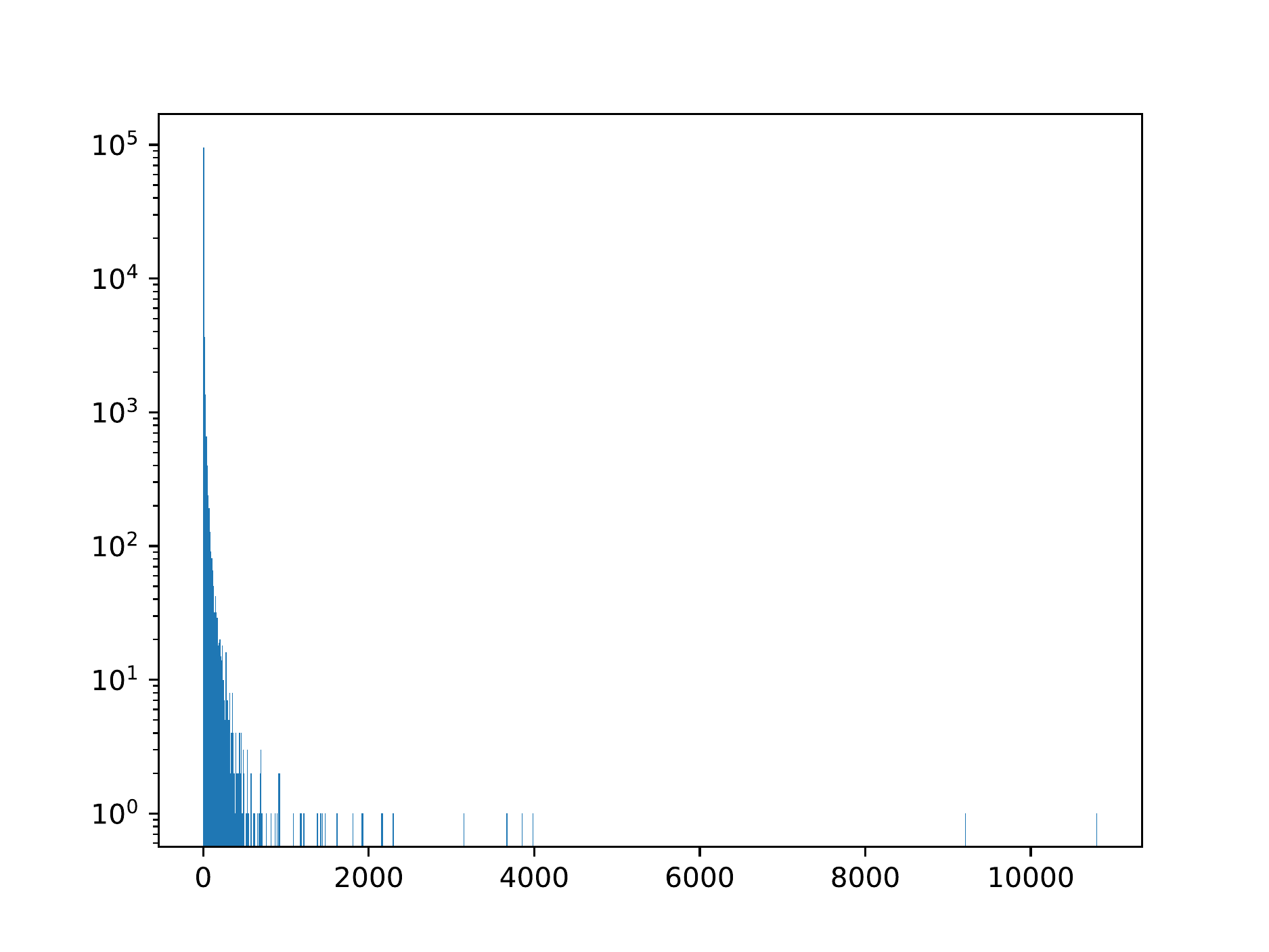} 
        \caption{\label{fig:1}Distribution of mention counts over the complete exploded CSM dataset.
         \textit{x}: distinct software mentions (\textit{log}), \textit{y}: sum of mentions for distinct software.}
    \end{minipage}\hfill
    \begin{minipage}{0.45\textwidth}
        \centering
        \includegraphics[width=\textwidth]{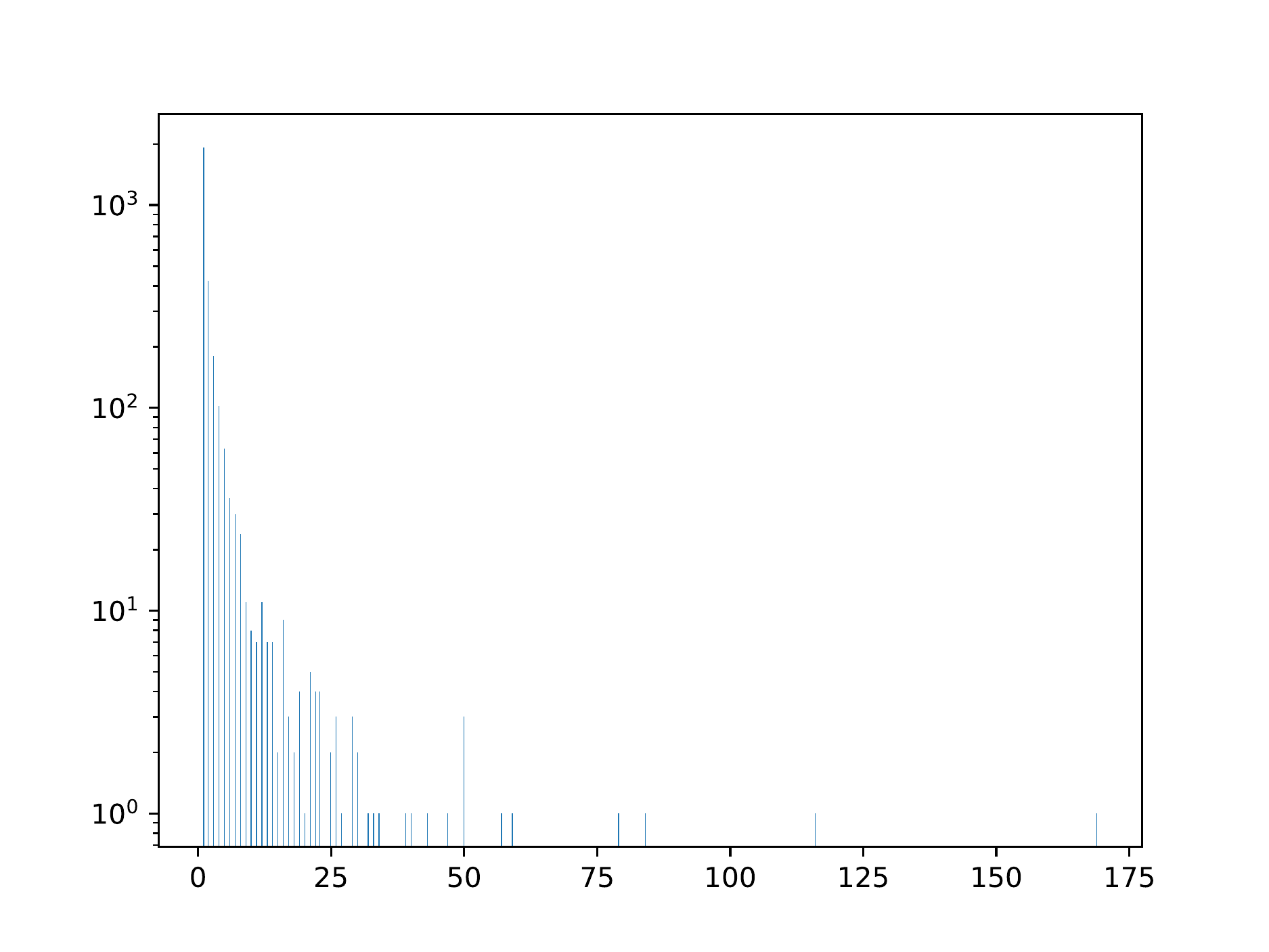} 
        \caption{\label{fig:2}Distribution of mention counts over our sample from the CSM dataset.
         \textit{x}: distinct software mentions (\textit{log}), \textit{y}: sum of mentions for distinct software.}
    \end{minipage}
\end{figure}

To obtain a sample from CSM for annotation, we first downloaded the dataset~\parencite{wadeCORD19SoftwareMentions2021a}.
It consists of a CSV file which collects publications in rows.
For each publication, there is some bibliometric information, such as DOI, title, source dataset, license, publication data, journal name and a list of URLs that resolve to a website providing a copy of the publication.
Additionally, the software mentions that were extracted from the publication are given as a comma-separated list, e.g., \texttt{['GraphPad Prism', 'SigmaPlot', 'Systat']}.
As we wanted to sample software mentions, not publications, we needed to explode the lists of software mentions in the dataset
while preserving the bibliometric information, and clean up the exploded mention strings.
The resulting data included $558,792$ software mentions.
To better understand the data, we took counts of software mentions and plotted their distribution (Figure~\ref{fig:1}).
The distribution shows extreme positive skew, around half the distinct mentinoned software has 1 mention, around 8\% has more than 10 mentions, less than 1.5\% have more than 50 mentions.
We then took a simple random sample of $1,000$ rows from the dataset, exploded it to get one distinct software mention per row, took counts of mentions in the sample and plotted their distribution (Figure~\ref{fig:2}).
The distribution of the sample also shows a strong positive skew.
As was to be expected, Levene’s test showed that the variances for number of mentions between the full dataset and our sample were not equal: $F(3296,35) = 7.73, p = .0054$.
As variance would not influence our study, we used this sample to take another random sample of 100 software mentions for developing and evaluating the annotation tagset.

CZI consists of several subsets~\parencite{IstrateEtAl2022a}, representing the different steps in the dataset creation process: raw data, linked data, disambiguated data.
As we were interested in evaluating the quality of original software mentions as found in the literature (RQ3), we used the \textit{raw} and \textit{linked} subsets.
To obtain a sample from CZI, we used two consecutive Jupyter notebooks.
In the first, we did the sampling.
In the second, we merged the sampled mentions with the \textit{linked} susbset of CZI to obtain
repository data for the mentions where available.

\begin{figure}[h!t]
    \centering
    \begin{minipage}{0.45\textwidth}
        \centering
        \includegraphics[width=\textwidth]{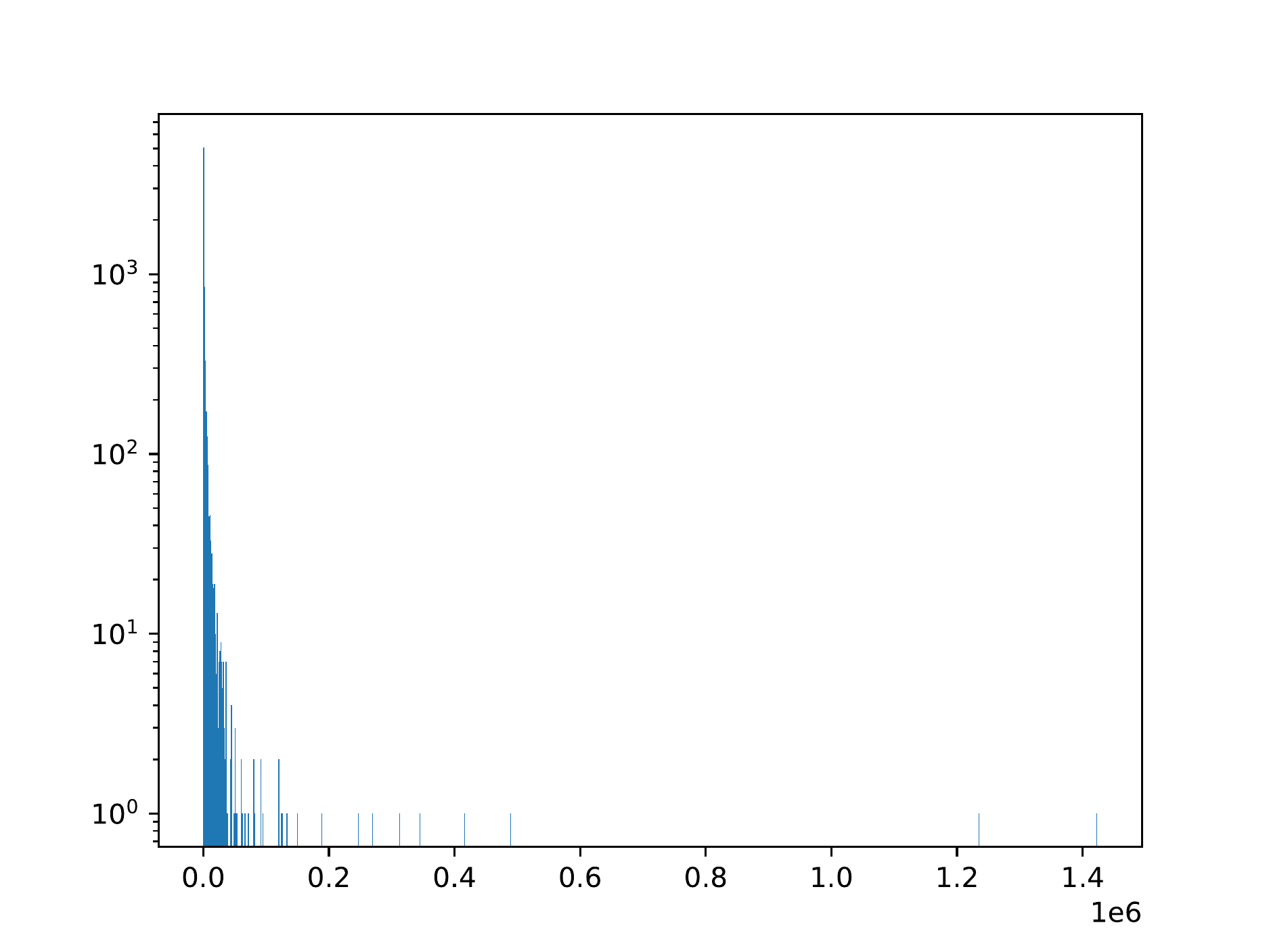} 
        \caption{\label{fig:3}Distribution of mention counts over the complete filtered CZI dataset.
         \textit{x}: distinct software mentions (\textit{log}), \textit{y}: sum of mentions for distinct software.}
    \end{minipage}\hfill
    \begin{minipage}{0.45\textwidth}
        \centering
        \includegraphics[width=\textwidth]{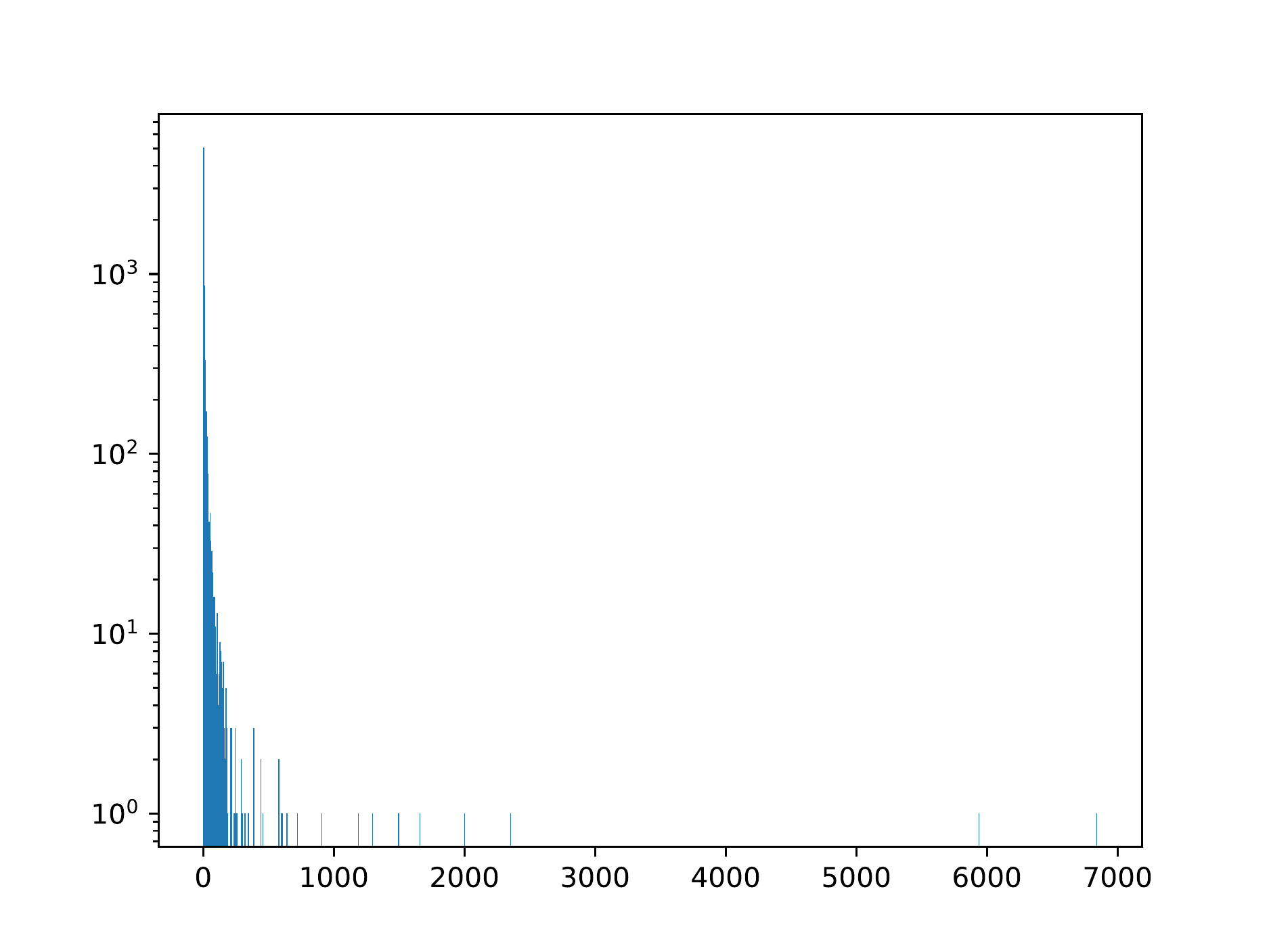} 
        \caption{\label{fig:4}Distribution of mention counts over our 100k sample from the CZI dataset.
         \textit{x}: distinct software mentions (\textit{log}), \textit{y}: sum of mentions for distinct software.}
    \end{minipage}
\end{figure}

As CZI is considerably larger than CSM, we used Dask~\parencite{TheDask2023.3.1developers2023} instead of Pandas~\parencite{Thepandasdevelopmentteam2023} for
working with the dataset, and ran the notebooks on a small CPU cluster at the German Aerospace Center's Institute for Software Technology using nbconvert~\parencite{Thenbconvert7.2.10developers2023}.
The sampling notebook downloaded and extracted the dataset archives.
We then merged the raw subsets for each of the collections contained in the CZI dataset (commercial, non-commercial, publishers collection),
and filtered those mentions that were curated as being to software (see~\textcite{IstrateEtAl2022a}).
The resulting dataset contained $20,792,352$ software mentions of $6,966$ distinct software.
As for CSM, we plotted the distribution of software mention counts in the dataset (Figure~\ref{fig:3})
As the dataset at this point still used $\approx1GB$ of disk space, 
we took a stratified proportionate sample of $\approx100,000$ rows from the dataset after brute-force deduplication of software names through capitalization,
to avoid long computation times.
The sample contained $99,973$ software mentions of $6,966$ distinct software.
To visually check the stratification, we plotted the distribution of software mention counts in the sample (Figure~\ref{fig:4}).
We then extracted one random row per distinct software name to avoid having duplicate instances of distinct software in our annotation sample,
and sampled 100 rows randomly for annotation.

CZI contains a subset with URLs resulting from exact searches for software names in different software source code and artifact repositories (see~\textcite[p. 8]{IstrateEtAl2022a}).
We wanted to leverage these data to add them to our assessment of the dataset quality (RQ1.1), and reuse the data to answer RQ2.
To achieve this, we merged the relevant information
from the subset of CZI containing repository data into
our sample. Identification of the relevant information
was possible through the unique mention IDs included in the raw and linked CZI subsets.

\subsection*{Annotation}
\label{subsec:methods:annotation}

We manually annotated the samples from CSM and CZI on different layers,
regarding the different research questions:
\begin{enumerate*}[label=(\alph*)]
    \item \textit{quality of the extracted mention}, e.g., whether the
    extracted string included the complete software mention string (RQ1);
    \item \textit{quality of the mention in the publication}, i.e.,
    whether and how the mention allows access to the software (RQ1);
    \item for CSM, a URL where the software could be found (RQ1);
    \item for CZI, \textit{quality of the repository links}, i.e.,
    whether links were extracted, and whether they referred to the correct software (RQ1);
    \item \textit{license and license type} for the mentioned software (RQ2);
    \item \textit{mention type} based on the mention types put forward in~\textcite[Table 6]{howisonSoftwareScientificLiterature2015} (RQ3).
\end{enumerate*}
Additionally, we added secondary annotations: whether the publication the mention was extracted from is a preprint; whether the publication the mention was extracted from is a paper describing the mentioned software; the confidence of the annotator regarding the correctness of the annotations.

\begin{figure}[h!t]
    \centering
        \resizebox{\textwidth}{!}{
        \begin{tikzpicture}[node distance=2cm]
            \node (dataset) [io] {CSM sample};
            \node (sampling) [process, below of=dataset] {Sampling};
        
            \node (annotation) [process, right of=sampling, xshift=2cm] {Subsample annotation};
            \node (confidence) [process, right of=annotation, xshift=2cm] {Confidence analysis};
            \node (confident) [decision, right of=confidence, xshift=2.5cm] {High annotation confidence?};
            \node (improvement) [process, below of=confident, xshift=-4.5cm] {Annotation \\ guidelines \\ improvement};
            \node (iaa) [io, right of=confident, xshift=2.5cm] {CSM sample 1 \\ ($n=50$)};
            \node (iaaannotation) [process, right of=iaa, xshift=2.5cm] {Annotation};
            \node (iaaout) [io, right of=iaaannotation, xshift=2.5cm] {Annotated CSM sample 1 \\ ($n=50$)};
            \node (agreement) [process, right of=iaaout, xshift=2.5cm] {Inter-annotator agreement};
            \node (finalannotation) [process, right of=agreement, xshift=2.5cm] {Annotation};
            \node (csm) [io, above of=finalannotation] {CSM sample 2 \\ ($n=100$)};
            \node (czi) [io, below of=finalannotation] {CZI sample \\ ($n=100$)};
            \node (analysis) [process, right of=finalannotation, xshift=2.5cm] {Analysis};
        
            \draw [arrow] (dataset) -- (sampling);
            \draw [arrow] (sampling) -- (annotation);
            \draw [arrow] (annotation) -- (confidence);
            \draw [arrow] (confidence) -- (confident);
            \draw [arrow] (confident.south west) |- node[anchor=north] {no} (improvement.east);
            \draw [arrow] (improvement) -| (annotation);
            \draw [arrow] (confident) -- node[anchor=south] {yes} (iaa);
            \draw [arrow] (iaa) -- (iaaannotation);
            \draw [arrow] (iaaannotation) -- (iaaout);
            \draw [arrow] (iaaout) -- (agreement);
            \draw [arrow] (agreement) -- (finalannotation);
            \draw [arrow] (csm) -- (finalannotation);
            \draw [arrow] (czi) -- (finalannotation);
            \draw [arrow] (finalannotation) -- (analysis);
        \end{tikzpicture}
        }
        \caption{\label{fig:workflow}Visualization of the complete assessment workflow.}
\end{figure}
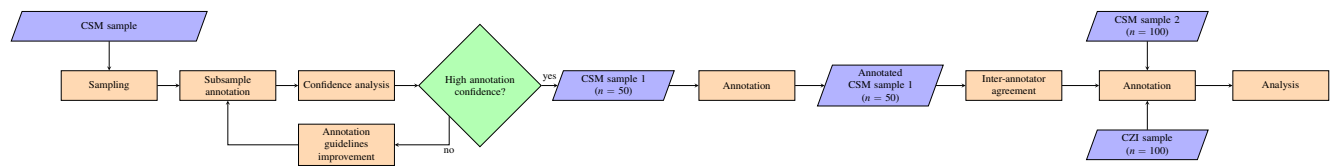

Annotations were guided by the annotation guidelines
summarized in the next section.
The guidelines were developed iteratively
by 
\begin{enumerate*}[label=\arabic*.]
    \item annotating random samples,\label{iter1}
    \item analysing confidence annotations,
    \item improving the annotation guidelines through discussion,
    \item repeating from~\ref{iter1} until the guidelines were not
    further improved.
\end{enumerate*}
Once no further improvements were made to the annotation guidelines,
SD, NCH, SB and OK each annotated the same set of 50 random mentions from the CSM sample.
We used these annotations to calculate inter-annotator agreement (Table~\ref{tab:iaa}).
\begin{table}[ht]
\centering
\begin{tabular}{ll}
\toprule
Annotation layer & Krippendorff's $\alpha$ \\
\midrule
Mention type	& 0.55 \\
Quality of the mention in the publication	& 0.72 \\
Quality of the mention extraction	& 0.65 \\
Preprint	& 0.80 \\
Software paper	& 0.49 \\
\midrule
All layers	& 0.64 \\
\bottomrule
\end{tabular}
\caption{\label{tab:iaa}Inter-annotator agreement for quality and mention annotations on a random sample of 50 mentions from the CSM sample.}
\end{table}
Finally, the annotation guidelines were applied to annotate software mentions
in a second sample from CSM ($n=100$) and a sample from CZI ($n=100$),
in addition to the already assessed annotations in the first CSM sample.
The complete workflow is shown in Figure~\ref{fig:workflow}.

\newpage

\subsubsection*{Annotation guidelines}
\label{subsubsec:methods:annotation:guidelines}

For each software mention in the sample,
\begin{enumerate}
    \item Resolve the first identifier for the publication in a web browser.
    \begin{enumerate}[label*=\arabic*.]
        \item If the publication is a preprint, use the next identifier if available.
        \item If the only available identifier is for a preprint,
        use the preprint.
    \end{enumerate}
    \item Open the PDF for the publication.
    \begin{enumerate}[label*=\arabic*.]
        \item If you cannot access the PDF due to a paywall,
        use the next identifier.
        \item If there is no next identifier, use Unpaywall\footnote{\url{https://unpaywall.org/}} to access
        an open version of the publication, or ask a co-author to retrieve the publication.
    \end{enumerate}
    \item Search for the exact mention string in the PDF.
    \item Verify for each search result that it is the exact search string. Note that:
    \begin{enumerate}[label*=\arabic*.]
        \item The mention string may be a substring of the complete software name (due to line breaks, composite names, etc.).
        \item There may be multiple software packages mentioned with similar names.
    \end{enumerate}
    \item Annotate the quality of the mention retrieval according to Table~\ref{tab:qa-retrieval}.
    \item Identify the best mention and annotate the mention type.
    \begin{enumerate}[label*=\arabic*.]
        \item Identify the best mention by adherence to the software citation principles.
        \item The \textit{Order} column in Table~\ref{tab:mention-types}
        encodes the quality of the mention (from 1 = best to 6 = worst)
        by principles:
        \begin{itemize}
            \item Importance is always the best. Citation of project name or website is better than citation of a publication. (Importance, Accessibility)
            \item Citation of a publication is better than citation of a user manual. (Credit)
            \item URLs in text are second best. (Accessibility)
            \item Instrument-like citation is better than name-only mention. (Accessibility)
            \item Name-only mentions are better than mention without name.
        \end{itemize}
        \item Only use mentions matching the exact mention string, including capitalization.
        \item Only URLs found in the same paragraph as the mention, or in a footnote that is called from the same paragraph, shall be annotated with \texttt{URL}.
        \item Citations to references must appear within the boundaries of the sentence that includes the mention.
        \begin{enumerate}[label*=\arabic*.]
            \item Examples for citations to process:
            \begin{itemize}
                \item ``We used SOFTWARE [1] for the analysis.''
                \item ``We used SOFTWARE for the analysis [1].''
                \item ``We used SOFTWARE for the analysis. [1]''
            \end{itemize}
            \item Example for citations to ignore:
            \begin{itemize}
                \item ``We used SOFTWARE and Otherthing for the analysis. We refuted the null hypothesis. The data provided evidence for something [1, 2].''
            \end{itemize}
        \end{enumerate}
    \end{enumerate}
    \item Annotate the quality of the mention (Table~\ref{tab:qa}).
    \begin{enumerate}[label*=\arabic*.]
        \item Differentiate between mention types \texttt{NA} and \texttt{SN}.
        \begin{enumerate}[label*=\arabic*.]
            \item If it is clear that the authors considered the mentioned entity software, annotate as \texttt{SN}.
            Examples: listed as ``computational method'', compared with other software.
            \item If still unclear, discuss with other annotators.
            \item If still unclear, annotate as \texttt{UN}.
        \end{enumerate}
    \end{enumerate}
    \item Annotate other layers.
\end{enumerate}

\begin{table}[h!t]
  \centering
    \begin{tabular}{ll}
    \toprule
    \textbf{Code} &	\textbf{Name} \\
    \midrule
    Y & Yes, name was correctly and completely retrieved from the publication for the dataset. \\
    N & No, name was NOT correctly and completely retrieved from the publication for the dataset. \\
    \bottomrule
    \end{tabular}%
  \caption{\label{tab:qa-retrieval}Annotations for quality of the mention extraction/retrieval.}
\end{table}%

\begin{table}[h!t]
  \centering
    \begin{tabularx}{\textwidth}{llXc}
    \toprule
    \textbf{Code} & \textbf{Name} & \textbf{Definition} & \textbf{Order} \\
    \midrule
    PUB & Cite to publication & Cites a paper/monograph primarily describing the mentioned software (NOT a review paper comparing different software), as it would for non-software cites.
    For non-software mentions, we don’t judge the suitability of the referenced work.
 &  2 \\
    \midrule
    PRO & Cite to project name or website & Cites the project name or website via a ``fake'' reference. & 1 \\
    \midrule
    URL & URL in text & URL in text or in footnote & 4 \\
    \midrule
    MAN & Cite to user manual &  & 3 \\
    \midrule
    INS & Instrument-like & Mention software in a manner similar to scientific instruments or materials, typically mentioning the name in text followed by the author or company and a location in parentheses. & 5 \\
    \midrule
    NAM & In-text name mention only &  & 6 \\
    \midrule
    NOT & Not even name mentioned &  & 7 \\
    \bottomrule
    \end{tabularx}%
  \caption{\label{tab:mention-types}Annotations for mention types following~\textcite{howisonSoftwareScientificLiterature2015}.}
\end{table}%

\begin{table}[h!t]
  \centering
    \begin{tabularx}{\textwidth}{lX}
    \toprule
    \textbf{Code} &	\textbf{Name} \\
    \midrule
    SC & Software where a direct link to a code repository or distribution repository landing page (e.g., CRAN, PyPI) can be found in the mentioning paper, and the page includes author\slash version\slash license metadata. \\
    \midrule
    SP & Software where a link to another website can be found in the mentioning paper and that website provides access to the source code, but the website does not provide author\slash version\slash license metadata. \\
    \midrule
    SN & Software but no link to a code repository or website providing access to the source code can be found in the mentioning paper. Annotate as SN even if the reference is to a software paper that does include a link to a source code repository. \\
    \midrule
    NA & Not software (only annotate this, retrieval quality and confidence) \\
    \midrule
    UN & Other classification - unknown\slash needs further investigation, e.g., unclear from the information in the paper whether this is software or not. \\
    \bottomrule
    \end{tabularx}%
  \caption{\label{tab:qa}Annotations for quality of the mention itself.}
\end{table}%

\subsubsection*{License annotation}
\label{subsubsec:methods:annotation:license}

Licenses for each software mention were annotated by NCH for the sample of CSM prepared during the initial hack event (see~\textit{\nameref{subsec:methods:datasets}}),
and by SD for the CZI sample.
This was done by examining any associated code repository, website or documentation related to the mentioned software. 
For CSM, where a link to a repository or project website was identified as part of the preliminary work of the hack event, this was used and checked to see if a license was documented.
If a link was not present in the initial dataset, an additional attempt to find a source of documentation for the software was undertaken by NCH, and the license recorded if available.
For CZI, any repositories for the software mention linked to from the \textit{linked} subset of CZI were checked to see if they were for the mentioned software.
If they were not, an attempt to find a source of documentation for the software was undertaken by SD and added to the dataset.
Additionally, the quality of the repository URL extraction in CZI was annotated, and the license recorded.
In a number of cases where the mention was to Software-as-a-Service and the license could not be identified, it was categorised in a subset of the "unknown" category.

\subsection*{Analysis}
\label{subsec:methods:analysis}

Results pertaining to research question \textit{RQ1: Are software mention datasets usable as data sources for research on research software?} and its subquestions
were gained through qualitative observation during the sampling and annotation work described above.

In order to yield exploratory results for research question \textit{RQ2: Is open source software more cited in a way that allows credit for software authors than closed source software?},
we grouped the licenses into the categories \textit{closed} (closed source licenses), \textit{academic} (academic use only, non-commercial licenses), \textit{permissive} (minimally restrictive open source licenses), \textit{copyleft} (open source licenses with reciprocal clauses) and \textit{unknown} (license conditions could not be found, including Software-as-a-Service).
We then clustered licenses into \textit{open} (\textit{permissive} + \textit{copyleft}) and \textit{closed} licenses (\textit{closed} + \textit{academic} + \textit{unknown}).
We also clustered mention types into quality categories \textit{good} (\texttt{PUB}), \textit{okay} (\texttt{PRO} + \texttt{URL}) and \textit{poor} (\texttt{INS} + \texttt{NAM}).

To gain exploratory insights into the data with regard to \textit{RQ3: Has the practice of software citation represented in software mention datasets improved in comparison to the practice described in~\textcite{howisonSoftwareScientificLiterature2015}?}, we grouped the mention type annotations by publication year of the mentioning publication, 
and compared the distribution over mention types
with the results from~\textcite{howisonSoftwareScientificLiterature2015}.

Analyses of the annotations were done in Jupyter Notebooks~\parencite{jupyter, Kluyver2016jupyter} 
using 
NumPy~\parencite{numpy, 2020NumPy-Array}, Pandas~\parencite{Thepandasdevelopmentteam2023} and Matplotlib~\parencite{matplotlib, Hunter:2007}.
Both our annotated data and the Jupyter notebooks with the analyses are available as part of~\textcite{druskatDonMentionIt2021}.

\section*{Results}
\label{sec:results}

We developed an approach to assess the usability of software mention datasets for research on research software (RQ1).
The approach includes taking a sample from the software mention dataset and preparing it for annotation,
then annotating it manually for mention extraction quality and mention categories and quality, following a set of annotation guidelines.
Finally, the annotations are analyzed to answer research questions.

\noindent We applied the approach to small samples (total $n=250$) from two software mention datasets (CORD-19 Software Mentions (CSM),~\textcite{IstrateEtAl2022b}; CZ Software Mentions (CZI),~\textcite{wadeCORD19SoftwareMentions2021a}),
and assessed it through qualitative observation.
We were particularly interested to see if software mention datasets can be used for
quantitative research that requires access to software metadata or artifacts (RQ1.1).
We also wanted to find out if they can be used for research into the practice of software citation (RQ1.2).

Through the application of our approach to assess the usability of software mention datasets for research on research software, we were able to uncover challenges to working with software mention datasets for the above-mentioned types of research.

We also found that practice of software citation must significantly improve in general 
to adhere to the software citation principles~\parencite{smithSoftwareCitationPrinciples2016},
thereby improving the quality of the source data used to create software mentions datasets.

\subsection*{Exploratory studies}
\label{subsec:results:exploratory-studies}

While it was our main goal to evaluate our approach to assessing the usability of software mention datasets for research on research software,
some highly preliminary results were gained during the evaluation process.

\begin{figure}[!ht]
    \begin{center}
        \noindent\resizebox{\textwidth}{!}{\input{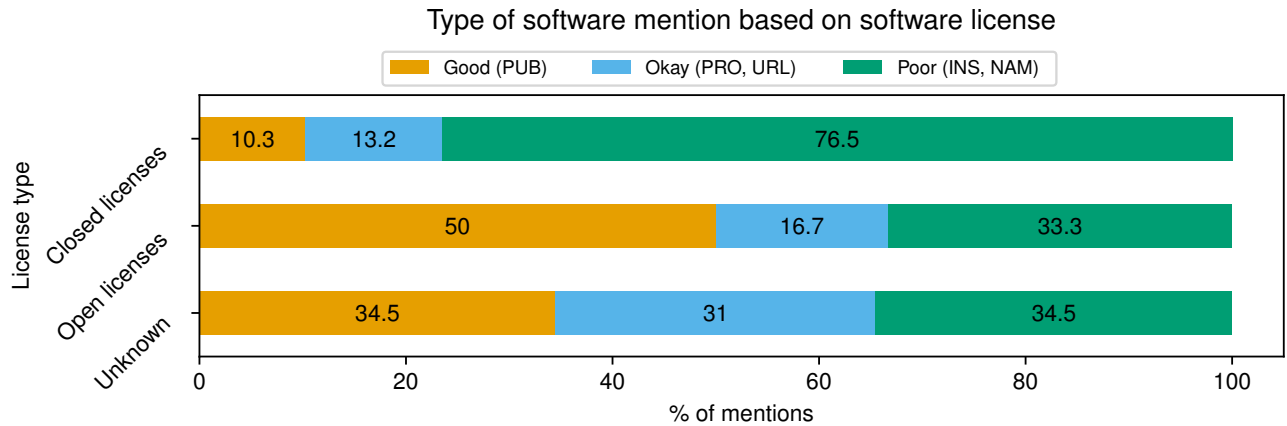}}
    \end{center}
    \caption{\label{fig:mention-type-vs-license}Percentages of mention types found in two samples from CSM and CZI, based on their software license, categorised by ``quality''.}
\end{figure}

Using our samples from CSM and CZI, we analyzed the software licenses for mentioned software.
Figure~\ref{fig:mention-type-vs-license} shows the distribution of license categories for closed and open licenses.


While software available under a closed license (license types ``closed'' and ``academic'', see Table~\ref{tab:software-license-categories}) is generally referenced using a poor quality mention (76\%),
half of the mentioned openly licensed software (``permissive'', ``copyleft'') is referenced using a good quality mention, and another 16.7\% using at least a medium quality mention.
Table~\ref{tab:mention-type-by-license} shows the detailed distribution of mention types over license types.

\begin{table}[ht]
\centering
\resizebox{\columnwidth}{!}{\begin{tabular}{rll}
\toprule
License category & License code & Description \\
\midrule
Closed     & Closed & Closed source licenses, generally commercial products \\
Academic   & Academic & No cost for academic or non-commercial use \\
Permissive & Apache, Artistic, BSD,  & Minimally restrictive open source licenses \\
{}         & MIT, Unlimited & \\
Copyleft   & LGPL, GPL & Open source licenses with reciprocal clauses \\
Unknown    & Unknown,  & License conditions could not be found, with a subset for \\
{}         & Unknown (SaaS) & Software as a Service (SaaS) with no license for service or code \\
\bottomrule
\end{tabular}}
\caption{\label{tab:software-license-categories}Categorisation of software licenses identified in our dataset.}
\end{table}

\begin{table}[ht]
\centering
\begin{tabular}{rcccccccccccc}
\toprule
{} & \multicolumn{2}{c}{PUB} & \multicolumn{2}{c}{PRO} & \multicolumn{2}{c}{INS} & \multicolumn{2}{c}{URL} & \multicolumn{2}{c}{NAM} & \multicolumn{2}{c}{Total} \\
{} &   $\sum$ &      \% &   $\sum$ &     \% &   $\sum$ &      \% &   $\sum$ &     \% &   $\sum$ &      \% &     $\sum$ &      \% \\
License type &     &        &     &       &     &        &     &       &     &        &       &        \\
\midrule
Closed       &   3 &   1.84 &   1 &  0.61 &  23 &  14.11 &   6 &  3.68 &  24 &  14.72 &  57.0 &  34.97 \\
Academic     &   4 &   2.45 &   0 &  0.00 &   1 &   0.61 &   2 &  1.23 &   4 &   2.45 &  11.0 &   6.75 \\
Permissive   &  16 &   9.82 &   2 &  1.23 &   0 &   0.00 &   4 &  2.45 &  12 &   7.36 &  34.0 &  20.86 \\
Copyleft     &  17 &  10.43 &   3 &  1.84 &   1 &   0.61 &   2 &  1.23 &   9 &   5.52 &  32.0 &  19.63 \\
Unknown      &  10 &   6.13 &   4 &  2.45 &   0 &   0.00 &   5 &  3.07 &  10 &   6.13 &  29.0 &  17.79 \\
\bottomrule
\end{tabular}
\caption{\label{tab:mention-type-by-license}Distribution of mention types over license categories.}
\end{table}


\noindent Using our samples from CSM and CZI, we analyzed the quality of software mentions published after 2015, 
and compared them with the mention quality reported in~\parencite{howisonSoftwareScientificLiterature2015}.
Mentions in publications published in or after
2016 made up $75\%$ of the CSM sample
and $63\%$ of the CZI sample.
Table~\ref{tab:mentionsdistribution} shows the distribution of mention types over the samples
from CSM, CZI and Howison and Bullard's data.

\begin{table}[ht]
\centering
\begin{tabular}{rcccccc}
\toprule
{} & \multicolumn{2}{c}{CZI} & \multicolumn{2}{c}{CSM} & \multicolumn{2}{c}{Howison \& Bullard} \\
Mention type &   $\sum$ &     \% &   $\sum$ &     \% &                 $\sum$ &     \% \\
\midrule
PUB &  19 &  30.2 &  20 &  30.3 &               105 &  37.2 \\
MAN &   1 &   1.6 &   0 &   0.0 &                 6 &   2.1 \\
PRO &   3 &   4.8 &   4 &   6.1 &                15 &   5.3 \\
INS &   4 &   6.3 &  11 &  16.7 &                53 &  18.8 \\
URL &   4 &   6.3 &  11 &  16.7 &                13 &   4.6 \\
NAM &  32 &  50.8 &  20 &  30.3 &                90 &  31.9 \\
\bottomrule
\end{tabular}
\caption{\label{tab:mentionsdistribution}Distribution of mention types over samples.}
\end{table}

\noindent Figure~\ref{fig:mention-type-comparison} shows the same data.
While the CSM data show an increase by 12.1\% for \texttt{URL} type mentions as compared to Howison and Bullard's data,
the other high and medium quality mentions have decreased (\texttt{PUB} $-6.9\%$) or only slightly increased (\texttt{PRO} $+0.8\%$).
For our CZI sample, the quality of mentions seems to have decreased overall: \texttt{PUB} $-7\%$, \texttt{PRO} $-0.5\%$.
There is only a slight increase in \texttt{URL} mentions ($+1.7\%$), but a significant increase in name-only mentions (\texttt{NAM}: $+18.9\%$),
which represent more than half of the mentions in the sample published after 2015.

\begin{figure}[!ht]
    \begin{center}
        \noindent\resizebox{\textwidth}{!}{\input{img/figure6.pgf}}
    \end{center}
    \caption{\label{fig:mention-type-comparison}Percentages of mention types found in the CSM and CZI samples compared to~\textcite{howisonSoftwareScientificLiterature2015}. See Table~\ref{tab:mention-types} for the definition of mention types.}
\end{figure}

\noindent When we clustered the mention types into three more coarsely grained categories,
we could observe the same trends. Table~\ref{tab:mentioncats} shows the detailed distribution,
Figure~\ref{fig:mention-cat-comparison} provides an overview.

\begin{table}[ht]
\centering
\begin{tabular}{rccc}
\toprule
{} & Good (PUB) & Okay (PRO, URL) & Poor (INS, NAM) \\
\midrule
CSM sample                 &       30.3 &            22.8 &            47.0 \\
CZI sample                 &       30.2 &            11.1 &            57.1 \\
Howison and Bullard (2015) &       37.2 &             9.9 &            50.7 \\
\bottomrule
\end{tabular}
\caption{\label{tab:mentioncats}Distribution of mention categories over samples.}
\end{table}

\begin{figure}[!ht]
    \begin{center}
        \noindent\resizebox{\textwidth}{!}{\input{img/figure7.pgf}}
    \end{center}
    \caption{\label{fig:mention-cat-comparison}Percentages of mention categories found in the CSM and CZI samples compared to~\textcite{howisonSoftwareScientificLiterature2015}.}
\end{figure}

\section*{Discussion}
\label{sec:discussion}

The application of our approach to two software mentions datasets allowed us to define features that software mention datasets should include 
to enable research into research software and software citation practices (see~\textit{\nameref{sec:results}}).

During sampling, we found that the structure of the dataset has an impact on ease of sampling.
CSM is based on publications: each row contains \textit{all} mentions from a specific publication in a Python-like string list.
This makes it more cumbersome and computationally expensive to extract a sample based on individual software mentions,
as all rows must first be exploded, and data must be cleaned afterwards to exclude artifacts from list data.
This structure also makes it harder to explore the dataset initially with regard to individual software mentions,
as they are potentially distributed across many rows, which makes filtering difficult.
This is much improved in CZI, which is based on individual mentions, starting with the initial raw data.
Linking the different subsets of CZI is straightforward, as mentions have a unique ID.
Therefore we reason that ordering data in software mention datasets by uniquely identified mention, not publication, is an important feature to make
the dataset easily usable.

As both datasets are tabular data, manual annotation is technically straightforward, as respective columns could be added.
Additionally, software for working with tabular data is readily available.
We reason that being persisted as tabular data is a feature that improves usability of datasets for research.

While the annotation process for mention types and quality, and the quality of mention extraction, towards answering RQ3 was generally straightforward,
the quality of the mention extraction has an impact on usability: in our sample of CSM, 19.3\% of mentions were incorrectly extracted,
i.e., the software name was not completely and correctly retrieved from the publication.
Our sample data from CZI suggests that this may have improved here, with 7\% of the mentions incorrectly extracted.
Generally, an analysis of mention types could be made much more feasible if the dataset included all contexts for all mentions of a software
across a publication. CZI includes context in its \texttt{text} column for raw data, which already makes it easier to find the respective mention.
However, a single context does not necessarily represent the ``best'' mention, nor does it allow for an analysis of how
a software is mentioned across a publication, where formal citations or references may have been made in another context.
We therefore reason that ideally, software mention datasets should include contexts for all mentions of a software across a publication.

Our samples included mentions that were not to software,
which may give a preliminary indication of the precision of the machine learning models that were used to extract the mentions.
In our CSM sample, extended with the sample used for calculating
inter-annotator agreement, 69 out of 150 mentions were not to software.
This means that only little more than half ($54\%$) of mentions were correctly identified as software.
In our CZI sample, 23 out of 100 mentions were not to software,
i.e. $77\%$ were correctly identified.
Assessing the precision of the models used for
software mention identification in more detail should be part of future research.
It seems necessary to improve these models to
achieve higher precision.

The annotation of license information for software mentions based on the two dataset samples -- as an example for
quantitative research on research software --
was difficult.
CSM does not contain any links between the mentioned software and its documentation or software repository.
This made it necessary to find a source of license documentation manually for each mentioned software.
Such an approach does not scale and renders quantitative research infeasible.
CZI, on the other hand, provide linking from different sources: GitHub, PyPI, CRAN, SciCrunch and Bioconductor (see~\textcite{IstrateEtAl2022a}).
However, three aspects impede quantitative research based on these data.
\begin{enumerate}
    \item CZI include links from different sources. In some cases, there were more than one link per software. This makes it difficult to
    leverage the links when they point to different targets. This is less problematic for cases where the targets represent the same software, 
    although this information is hard to assess automatically. It becomes highly problematic, when the targets represent different software,
    which additionally may have the same name. In our sample of CZI, 7 out of 62 mentions (11.3\%) had links to multiple different software.
    \item Linking was created automatically in CZI, but the linking quality is generally poor: out of 55 mentions in our sample for which one,
    or multiple of the same link were provided, 36 (65.4\%) gave a link to the wrong software. Of these, 9 were based on exact string matches between
    software name and mention string.
    \item Not all software mentions have a link associated with them (16 out of 78 potentially linkable software mentions, 20.5\% in our sample)
\end{enumerate}
We are hopeful that the quality of machine learning models and algorithms for linking software mentions to repositories will be improved in the future.
We nevertheless reason that links to repositories are not currently a dataset feature that improves the usability of mention datasets for quantitative research.
Alternatively, datasets with repository URLs mined directly from publications (e.g.,~\textcite{EscamillaEtAl2022a}) could be used for quantitative research,
although the semantics between the contents of the publication and the URL cannot be established to any satisfactory degree.
One approach to improve the latter would be to combine URL mining and mention retrieval in the same dataset.

The underlying issue in terms of the challenges of using software mention datasets for research purposes as discussed here
is the currently still suboptimal practice of software citation.
If authors more strictly followed the software citation principles~\parencite{smithSoftwareCitationPrinciples2016},
software mentioned in the literature would be accessible per default, not via tedious manual search.

We believe that our approach to assess the usability of software mention datasets is generally valid,
but it comes with its own shortcomings.
Manual annotation of samples does not scale, and thus our approach can only be used for preliminary results
from exploratory studies.
The approach could potentially benefit not only from better linking models and algorithms as mentioned above,
but also from a machine learning model to categorize mention types.
Another shortcoming of our approach is the use of mention types that do not optimally reflect the
software citation principles. Instead, we reused the types from~\textcite{howisonSoftwareScientificLiterature2015}
to achieve comparability.
Future work could therefore attempt the development of new mention type categories closed related to the software citation principles.

Based on the exploratory research presented here, we recommend features that software mention datasets can include to enable research on research software:

\begin{enumerate}
    \item Software mentions datasets should be ordered by uniquely identified mention, not by mentioning publication.
    \item Software mentions datasets should be made available as tabular data.
    \item Software mentions datasets should include contexts for all mentions of a software across a publication.
    \item Software mentions datasets should only include mentions that are to software, not to other entities.
    \item URLs to software repositories in software mentions datasets should resolve to repositories containing the mentioned software.
\end{enumerate}

\subsection*{Exploratory studies}
\label{subsec:discussion:exploratory}

The results from our exploratory studies (see~\textit{\nameref{subsec:results:exploratory-studies}}) are not representative.
Their following discussion is therefore highly preliminary.

For \textit{RQ2: Is open source software more cited in a way that allows credit for software authors than closed source software?}
we hypothesized that commercial\slash close source software is cited more frequently using a lower-quality in-text name mention or citation to project name or website, and that open source software 
is cited more frequently with a repository or associated research publication in the reference.
The results from our sample annotations seem to support this hypothesis.
They showed that a third of openly licensed software still has a poor quality mention in our sample, suggesting that efforts towards better software citation are still necessary.

We were also interested in finding out the state of software citation -- or mentioning -- as compared to 2015.
For \textit{RQ3: Has the practice of software citation represented in software mention datasets improved in comparison to the practice described in~\textcite{howisonSoftwareScientificLiterature2015}?}
we hypothesized that categories of mentions that
reflect the principles better are found relatively more often in 
mentions from publications in the dataset samples that were published in or after 2016,
than in the data presented in~\textcite{howisonSoftwareScientificLiterature2015}.
Our results suggest that -- at least in our data samples -- the practice of software citation has not improved in the last 8 years.

The results from both exploratory studies would support previous work that describes challenges for software citation~\parencite{katz2019implementationchallenges,jaySoftwareMustBe2021}
and call for improved practice of software citation~\parencite{Bouquin2023report} and advocacy towards it~\parencite{DuEtAl2022a}.

\subsection*{Limitations}
\label{subsec:threats}

The positive evaluation of our assessment approach is threatened by medium value Krippendorff's $\alpha$ inter-annotator agreement scores (Table~\ref{tab:iaa}),
where a value of $\alpha 1.0$ signifies complete agreement.
Regarding the central ``mention type'' annotation, agreement showed only a slim positive trend ($\alpha 0.55$), and $\alpha 0.64$ across all layers.
More confidence can be put into the assessment of the similarly important mention extraction quality annotations, at $\alpha 0.72$.
Likewise, the qualitative observation towards answering RQ1 (RQ1.1, RQ1.2) is subjective by nature.

Our studies towards answering RQ2 and RQ3 were based on very small sample sizes.
This would be a severe threat to validity of results that claim more than exploratory significance; note that we do not make such a claim here.
Additionally, both sampled datasets include mostly biomedical software, and their data could not be used to make claims for research software in general.

\section*{Conclusion}
\label{sec:conclusions}

We presented an approach for the assessment of the usability of software mentions datasets for research on research software.
Despite some small shortcomings, our approach is valid and applicable in the assessment of software mention datasets.
It includes sampling and data preparation, manual annotation for quality and mention characteristics, and annotation analysis.
We applied our approach to two software mention datasets (\textcite{IstrateEtAl2022b,wadeCORD19SoftwareMentions2021a}) to evaluate the approach.
Through our approach, we were able to find challenges to working with the selected datasets.
We were also able to define dataset features that would enable the use of software mention datasets in quantitative research on research software,
and research into the practice of software citation.
These features include: a dataset structure based on individual software mentions, persistence as tabular data, the inclusion of contexts for all mentions of individual software across an individual publication.
We also found that automatically retrieved links to repositories for a software were of low quality in our sample;
better retrieval algorithms and machine learning models are needed to improve the quality of links.
Finally, the underlying issue with and challenge to working with software mention datasets
is the suboptimal practice of software citation.
We conclude that software should be \textit{cited} using a formal citation and reference according to the software citation principles~\parencite{smithSoftwareCitationPrinciples2016},
and \textit{not mentioned}.

\section*{Acknowledgments}

This work originated from an idea proposed by NCH and SD at the Software Sustainability Institute's Collaborations Workshop 2021 and we acknowledge feedback from Michelle Barker, Daniel S. Katz, Shoaib Sufi, Carina Haupt and Callum Rollo. Our immense thanks to the other members of the team who explored the idea during the hackathon: Hao Ye, Louise Chisholm and Mark Turner.

We thank two anonymous reviewers of the original submission for constructive comments that helped shape this revised version.

\printbibliography

\end{document}